\documentclass[reprint,aip,amsmath,amssymb]{revtex4-2}

\usepackage{graphicx}
\usepackage{dcolumn}
\usepackage{bm}

\usepackage[utf8]{inputenc}
\usepackage[T1]{fontenc}
\usepackage{mathptmx}
\usepackage{etoolbox}
\usepackage{xcolor}
\usepackage{braket}
\usepackage{float}
\usepackage[english]{babel}
\makeatletter
\def\@email#1#2{%
 \endgroup
 \patchcmd{\titleblock@produce}
  {\frontmatter@RRAPformat}
  {\frontmatter@RRAPformat{\produce@RRAP{*#1\href{mailto:#2}{#2}}}\frontmatter@RRAPformat}
  {}{}
}%
\makeatother
\begin{document}

\preprint{AIP/123-QED}

\title{Integration of hBN Single-Photon Emitters into a Hybrid Optomechanical Membrane-in-the-Middle Fiber-Cavity 
}
\author{P. Maier}
\author{A. Kubanek}
\email{alexander.kubanek@uni-ulm.de}
\affiliation{Institute for Quantum Optics, Ulm University, Albert-Einstein-Allee 11, 89081 Ulm, Germany.}

\date{\today}

\begin{abstract}
The integration of membranes into optical resonators plays a key role in a variety of applications, including optomechanics. Membranes hosting single photon emitters, ideally with access to spin states, open new roads in optomechanics, spin-mechanics and spin-optomechanics. Hexagonal boron nitride is among the most promising two-dimensional materials showing excellent optical and mechanical properties combined with the ability to host optical active (spin-) defects. The deterministic creation of optically active defect centers in hexagonal boron nitride membranes and their coupling to optomechanical systems is an outstanding challenge. Here, we explore an alternative, hybrid approach in order to establish coupling between a single photon emitter in commercially available hexagonal boron nitride flakes and a fiber-cavity mode. We solve technical challenges, such as scattering losses arising from an uncontrolled flake topography, and establish deterministic hexagonal boron nitride positioning on the cavity mirror. For the coupled system we observe cavity induced spectral enhancement by a factor of up to 100 at room temperature. We extend our work by positioning the single photon emitter in hexagonal boron nitride flakes on a highly-strained silicon nitride membrane in a membrane-in-the-middle configuration. The mechanical vibrational modes of the silicon nitride membrane and cavity-coupled emission from the single photon emitter are simultaneously observed. Our work is a first step towards the realization of a cavity optomechanics platform with an incorporated single photon emitter and provides a starting point to explore hybrid spin–optomechanics. 

\end{abstract}

\maketitle

\section{Introduction}

\begin{figure*}
    \centering
    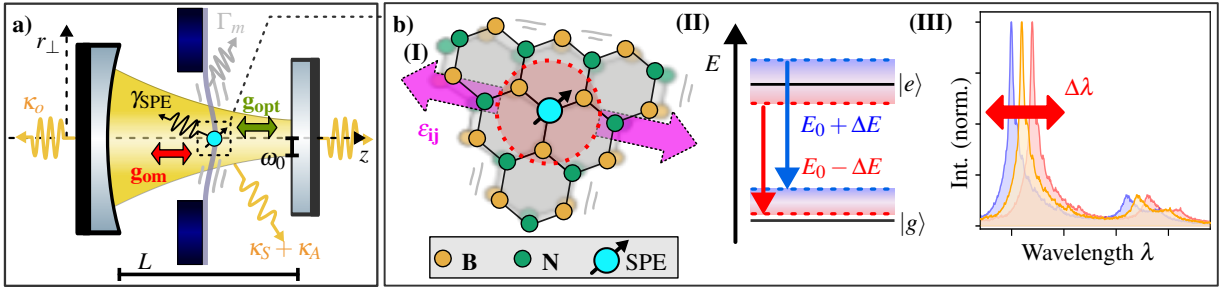
    \caption{Schematic overview of an an optomechanical system with an incorporated strain sensitive SPE in hBN. \textbf{a)} Illustration of an open Fabry-Perot microcavity and an optomechanical membrane (membrane-in-the-middle configuration). A photon interacts with a phonon of the membrane with the coupling strength of $g_{\text{om}}$, while a photon emitted from a SPE with the rate $\gamma_{\text{SPE}}$ couples with the rate $g_{\text{opt}}$ to the optical mode of the cavity. Dissipative effects for the mechanical and optical resonators are represented by the damping rates $\Gamma_m$ and $\kappa = \kappa_o + \kappa_S + \kappa_A$, where $\kappa_o$ is the outcoupling rate through the mirrors, $\kappa_S$ and $\kappa_A$ are scattering and absorption rates, respectively. \textbf{b)} Schematic representation of a SPE in hBN under the influence of phonons and strain. \textbf{(I)} Schematic representation of a spin defect inside a hBN monolayer. Application of strain described by a tensor $\epsilon_{\text{ij}}$ changes the energy-level structure \textbf{(II)}. The atomic transition energy $E_0$ is changed by $\Delta E$ resulting in a shift of the emission wavelength $\lambda$ \textbf{(III)} of the defect. Combined with a configuration as described in a) this platform could realize an interface between photons to phonons and phonons to optical (spin-)defects.}
    \label{Optomechanics Sketch}
\end{figure*}

Optomechanics investigates the coupling of optical and mechanical degrees of freedom based on the fundamental principle that light carries momentum and exerts radiation pressure on matter as well as coupling mediated by dipole forces  \cite{Frisch1933,Kippenberg2008,Aspelmeyer2012,Aspelmeyer2014,Barzanjeh2022} . 
The interaction strength between light and mechanical motion can be enhanced by means of optical cavities \cite{Cohadon1999}. Cavity-optomechanics with macroscopic objects, such as a movable cavity mirror, enables for example, entanglement on macroscopic scale \cite{Vitali2007,Riedinger2018}. Miniaturization of cavity-optomechanics is attractive leading to increasing coupling rates between photons and mechanical motion while sustaining optical and mechanical quality factors \cite{Groeblacher2009}. Prominent platforms include microdisk resonators \cite{Mitchell2016,Shandilya2019,Joe2024}, optomechanical crystals \cite{Eichenfield2009,GomisBresco2014,Burek2016}, microwave cavities \cite{Regal2008,Teufel2008,Teufel2011,Bothner2020,Seis2022,Bozkurt2023}, levitated systems \cite{Chang2009,Neukirch2015,Pontin2018,Delic2020,GonzalezBallestero2021,Millen2020,Millen2020a,Stickler2021} and open Fabry-Perot cavity based experiments \cite{Purdy2010,Ruelle2022a,SanchezArribas2023,Tenbrake2024}, where records in mechanical quality factors above $10^9$ for a Si$_3$N$_4$ membrane inside an optical cavity have been reported \cite{Rossi2018}.\\
Coupling additional degrees of freedom of a quantum object, such as an atomic transition, to an optomechanical systems provides access to complex quantum dynamics based on hybridizing radically different quantum systems \cite{Arcizet2011}.
Empowering cavity quantum electrodynamics in cavity optomechanics, offers new paths for high-fidelity quantum state control, coherent quantum transduction or mediated spin control \cite{Rabl2009, Abdi2019}. Furthermore, deterministic emission of single photons paves the way to study back action of single photon states in a hybrid approach. Applications include frequency combs with frequency spacings as small as the fundamental vibrational mode \cite{Abdi2019}. \\
To explore optomechanics with integrated single photon emitters (SPEs) two criteria need to be fulfilled. First, membranes with excellent optical and mechanical properties are required. Second, the same membrane needs to host SPEs. A promising candidate is hexagonal boron nitride (hBN), a two-dimensional material with a large bandgap of 6$\,$eV, transparent in the optical and infrared spectrum \cite{Cassabois2016}, low photothermal heating \cite{Shandilya2019, SanchezArribas2023, Zheng2017, Falin2017, Linderaelv2021}, low effective mass $m_{\text{eff}}$, high Young's modulus \cite{Zheng2017} and a high breaking strain \cite{Falin2017}. 
First hBN-based cavity optomechanics experiments have been mechanically exfoliated hBN-resonators coupled to the near-field of microdisk cavities \cite{Shandilya2019,Liu2021}. Shandilya et al. \cite{Shandilya2019} demonstrate coupling of the thermal motion of the hBN-resonator through its optomechanical interaction with a silicon microdisk cavity, with a sensitivity of 0.16 pm/$\sqrt{Hz}$. hBN-membranes integrated in membrane-in-the-middle optomechanics were used to demonstrate radiation pressure backaction \cite{SanchezArribas2023}. hBN is also known to host optically active defect centers \cite{Cholsuk2024}. A variety of SPEs in hBN \cite{Hoese2022,Zeng2022, Samaner2022,AlJuboori2023,Bourrellier2016,Martinez2016,Tran2016,Grosso2017,Exarhos2019,Dietrich2020,Fournier2023, Koch2024} have been investigated, some of which have shown sensitivity to applied mechanical strain \cite{Mendelson2020, Shaik2022} or electric fields \cite{Li2025}, or optically addressable spin states \cite{Froech2021,Liu2022,Gottscholl2021,Gao2022,Stern2022,Guo2023,Stern2024, Wu2025}. Creation, engineering and tuning of defect centers in hBN has been investigated \cite{Bourrellier2016,Mendelson2020, Mendelson2021,Ziegler2019,Vogl2018,Chen2021,Gao2021,Kumar2023, Kianinia2020,Fournier2021,Nonahal2023,Chen2021a}. However, no cavity-optomechanical system hosting a SPE in hBN has been reported, partly because deterministic creation of defect centers in hBN membranes remains challenging. \\
Here, we propose an alternative route by using SPEs in commercially available hBN flakes. In contrast to thin membranes \cite{Froech2020,Haeussler2021,Vogl2019}, the hBN flakes naturally host optically-active SPEs, but their integration into optical cavities is challenging. The precise location of the SPE within the hBN flake is uncontrolled and requires precise alignment methods. Also, the uncontrolled topography of the hBN flakes leads to significant scattering losses. Scattering losses hinder not only the construction of optomechanical devices, but also prevent to increase optical efficiency and to establish cavity funneling of spectrally narrow lines \cite{Albrecht2013, Vogl2019,Froech2020,Haeussler2021, Kuruma2021, Froech2022}. We overcome issues arising from scattering losses by developing a toolset of manipulation and transfer techniques improving the flakes' topographies. We demonstrate overall integrability of these flakes into open Fabry-Perot fiber cavities (FPFC), observing strongly enhanced emission rate and reduced linewidth when coupling the atomic transition to the optical mode. We further explore the potential as optomechanical system coupled to a SPE. Therefore, we propose a hybrid solution, based on a high-stress silicon nitride (Si$_3$N$_4$) membrane-in-the-middle (MiM) cavity configuration on which we deterministically place hBN flakes with a topography exhibiting reduced scattering. We observe distinct mechanical modes from the Si$_3$N$_4$-membrane which persist after hBN placement. The hBN flakes host precharacterized SPEs which couple to the mode of the MiM cavity system.

\section{A hybrid approach towards optomechanical membrane-in-the-middle cavity system with integrated SPE}

\begin{figure*}
\centering
\includegraphics[scale = 1]{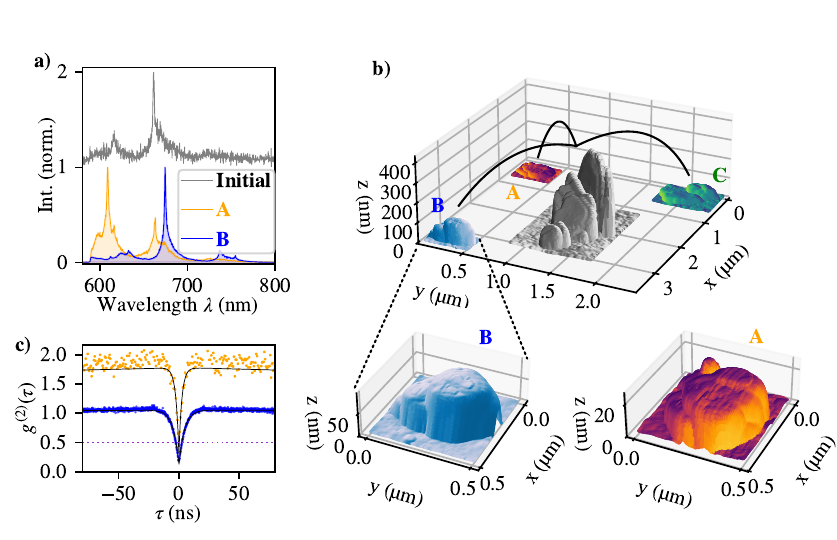}
\caption{Exemplary extraction process of hBN structures hosting single photon emitters (SPEs). \textbf{a)} Emission spectra of the initial ensemble of hBN particles (grey) and of the extracted flake structures A (orange) and B (blue). \textbf{b)} Topography of the initial ensemble (grey) and flake structures after extraction efforts with an AFM (blue, orange and green) and corresponding zoomed-in visualizations of flake structures A and B. \textbf{c)} Second-order autocorrelation measurements for emitters A and B and corresponding fits, confirming single photon emission with $g^{(2)}_A(0)= 0.32$ and $g^{(2)}_B(0)= 0.13$. The bunching behavior observed for emitter A is attributed to a metastable shelving state.}
\label{DeClustering}
\end{figure*}

Integrating a hBN membrane with incorporated SPEs into the mode of an optical cavity opens up new opportunities to establish multiple coupling mechanisms in one system.\\
First, the radiation-pressure interaction between optical photons and the mechanical motion of the membrane. The coupling rate is described by

\begin{equation}
g_{\text{om}} = G \cdot z_{\text{ZPF}} = \frac{\delta \omega_{c}}{\delta z} \cdot \sqrt{\frac{\hbar}{2m_{\text{eff}} \Omega_m}}
\end{equation}

where $z_{\text{ZPF}}$ denotes the zero point fluctuation of the mechanical oscillator, $\omega_c$ the optical frequency of the cavity mode, $m_{\text{eff}}$ the effective mass and $\Omega_m$ the mechanical eigenfrequency. For a MiM configuration in a Fabry–Pérot cavity, the frequency pulling factor $G$ is determined by the membrane reflectivity $r$ and its position $z$ within the intracavity standing wave by 

\begin{equation}
G=\frac{\delta\omega_c}{\delta z} \overset{|r|\ll 1}{\approx} 2\frac{\omega_c}{L}\lvert r \rvert \sin \left(2\frac{\omega_c}{c} z \right)
\end{equation}

where $L$ describes the length of the optical cavity under the assumption of weak membrane reflectivity ($r\ll 1$).\cite{Jayich2008,Thompson2008,Aspelmeyer2014,Saarinen2023} A high coupling rate is achieved when miniaturizing the membrane towards low effective mass $m_{\text{eff}}$. Applied to our proposed system, we estimate that with a tailored few layer hBN flake with lateral dimension large enough to avoid significant optical clipping losses (e.g. $d_{\text{hBN}}\approx10\,$µm), the effective mass $m_{\text{eff}}$ of a vibration mode at $\Omega_m \approx 1\,$MHz will be in the range of a few femtograms (similar to previously reported values \cite{SanchezArribas2023}). At the same time, mechanical and optical quality factors $Q_m = \frac{\Omega_m}{\Gamma_m}$ and $Q_o = \frac{\omega_c}{\kappa}$ need to be maintained, where $\Gamma_m$ and $\kappa$ are the mechanical and optical dissipation rates of the system (as represented in fig. \ref{Optomechanics Sketch}a)). Favorable materials provide a high Young's modulus $E$ (allowing high mechanical frequencies $\Omega_m$). Also, the materials must have low optical dissipation rates $\kappa_A + \kappa_S$, provided by low absorption rates and smooth surface topographies, eliminating scattering losses. \\
Second, the optical coupling strength between the SPE and the cavity mode is position dependent and therefore sensitive to the motion of the SPE together with its host membrane. The out-of-plane stiffness of hBN is very small leading to a large zero-point amplitude and therefore a high sensitivity to the motion. The optical coupling of the SPE to the cavity mode is thereby described by 

\begin{equation}
g_{\text{opt}}(z,r_\perp) = \sqrt{\frac{3\lambda^2 c \gamma_{\text{SPE}}}{4\pi V_m}} \cdot \xi(z,r_\perp)
\end{equation}

where $\gamma_{\text{SPE}}$ is the natural linewidth of the emitter \cite{Hunger2010, Kaupp2016}. The factor 

\begin{equation}
\xi(z,r_\perp) = \cos\left(2\frac{\omega_c}{c} z\right) \exp \left( \frac{-r_\perp^2}{\omega_o^2}\right)
\end{equation}

is a position dependent coupling factor, accounting for the overlap between the dipole moment of the SPE and the electric intracavity field. The coupling rate $g_{\text{opt}}$ is modulated along $z$ due to the positioning of the SPE in the standing wave and in the lateral direction $r_\perp$ due to the positioning regarding the cavity mode waist radius $\omega_0$ (assuming a TEM$_{00}$ Gaussian mode).\cite{Colombe2007} \\
Third, the optical transition of the SPE couples to the mechanical motion of the membrane via strain (as depicted in fig. \ref{Optomechanics Sketch}b)). The perturbation of energy levels is described by the Hamiltonian $\hat{H}_{\text{strain}} = \sum_{ij} \hat{\sigma}_{ij}\,\epsilon_{ij}$ with the strain tensors $\epsilon_{ij}$ and the orbital operators $\hat{\sigma}_{ij}$. \cite{Grosso2017} The first order change in the atomic transition energy $E_0$ is given by

\begin{equation}
\Delta E \approx \epsilon \left( \kappa_{xx} \cos^2 \Phi + \kappa_{yy} \sin^2 \Phi \right)
\end{equation}

where $\Phi$ describes the angle between the defects dipole moment and the strain axis, $\epsilon$ the strain magnitude and $\kappa_{xx}, \kappa_{yy}$ the strain susceptibility parameters of the optical defect. It is noted that out of plane strain ($\epsilon_{zz}\approx 0)$ caused by the Poisson effect is assumed to be negligible 
\cite{Mendelson2020}. Defects in hBN membranes with high susceptibilities to strain have been reported \cite{Grosso2017,Mendelson2020,Cholsuk2022,Shaik2022} enabling an estimated sensitivity to the mechanical motion in out-of-plane direction $z$ of up to $\frac{\delta \omega(z)}{z^2} \approx  60\,\frac{\text{MHz}}{\text{nm}^2}$. The coupling could be resolvable using SPEs with narrow spectral lines \cite{Dietrich2018,Hoese2020} in a MiM configuration.\\
In this work, we propose a hybrid approach leveraging SPEs in hBN flakes and high-stress Si$_3$N$_4$-membranes \cite{FlowersJacobs2012, Rochau2021}. The platform could provide reasonable high interaction rates for optical coupling to the mechanical mode as well as mechanical coupling to the atomic transition. However, due to the hybrid approach the hBN flakes are transferred onto the Si$_3$N$_4$-membrane and the strain needs to be mediated through the hBN-Si$_3$N$_4$ interface. Therefore, an open question remains regarding the transfer of mechanical strain through the bonding interface between the hBN flake and the Si$_3$N$_4$-membrane.\\
Further potential arises for coupling to spin states leading to additionally spin-mechanical degrees of freedom.  A description for the interaction of spin states with phonons is given e.g. by the Huang-Rhys model \cite{Doherty2013, Exarhos2017, Feldman2019}. The spin-mechanical system can be engineered to optimize spin-motion interaction, capable for initialization, rotation, and readout of the SPEs spin qubit \citep{Abdi2017}. Alternatively, magnetic coupling between the mechanical oscillator and the SPE's spin can be induced by immersing the system in a strong magnetic field gradient resulting in spin-dependent forces \citep{Arcizet2011}.

\section{Manipulation and preparation of individual defects in \NoCaseChange{h}BN flakes}

\label{AFM_Decluster}

Suitable SPEs are pre-selected with a confocal microscope and extracted from clusters with an AFM or a tungsten-tip-based manipulation system. As a first step, a source sample of hBN flakes is prepared by spin-coating or drop-casting a commercially available hBN emulsion (2D Semiconductors) onto a fused silica substrate following ultrasonic bath treatment. The spin-coated sample is annealed under vacuum to improve the optical properties of optically active emitters \cite{Hoese2020}. An exemplary SPE spectrum is represented in fig. \ref{DeClustering}a) (grey). To investigate the topographic properties of the pre-selected emitters host, we conduct AFM scans. As presented exemplary in fig. \ref{DeClustering}b) (grey) the emitter is embedded in an ensemble of a multitude of individual flakes. It is speculated that only one flake is hosting the emitter (excluding the possibility of the emitter being hosted between two or more particles). We therefore dismantle ensembles into sub-ensembles and separate them by a few micrometers (fig. \ref{DeClustering} (blue, orange and green)) with an AFM by driving the cantilever into contact with the individual flakes. The sub-ensembles are investigated again for their optical properties (as shown exemplary for flake structures A and B in fig. \ref{DeClustering} c)). With a thickness below $27\,$nm (A) (less than a tenth of the operation wavelength) and 100$\,$nm (B) and lateral extension in the micrometer range they are particularly interesting for the integration into scattering-sensitive micro-cavities. The flake structures hosting emitters can then be transferred to a target sample as described in supplementary section I. Particles without SPEs can be discarded (e.g. structure C in fig. \ref{DeClustering} b)) and remain on the source substrate.

\begin{figure*}
\centering
\includegraphics[scale = 1]{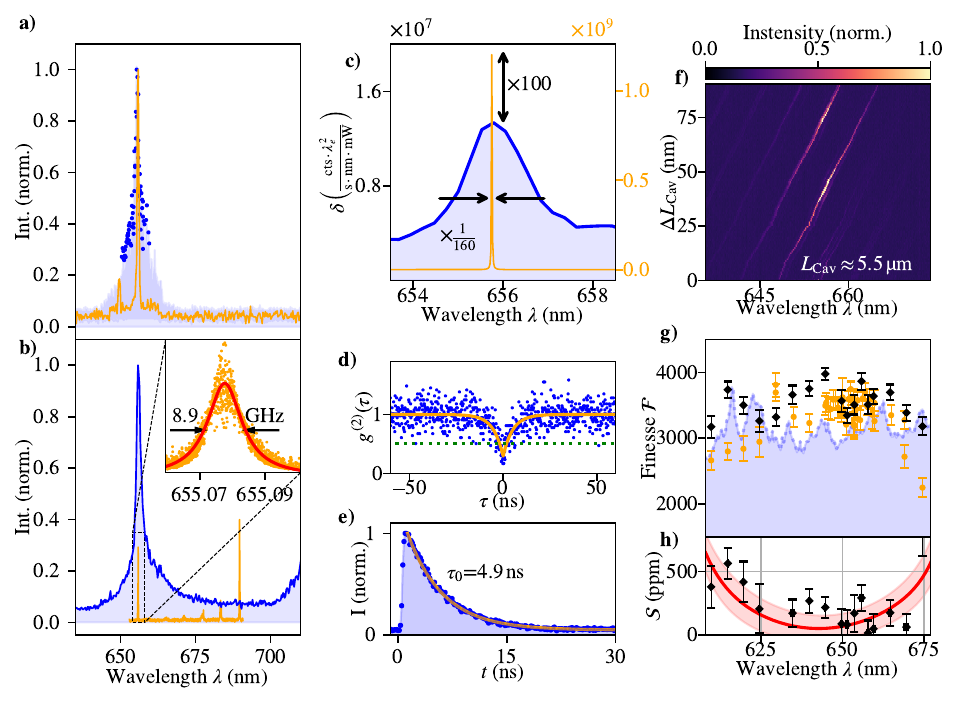}
\caption{FPFC coupled SPE in a transferred and manipulated hBN flake. \textbf{a)} Normalized coupled emitter-cavity spectrum (orange) of a SPE and cavity probed free-space emission spectrum (blue) when exciting with a $\lambda=532\,$nm laser. \textbf{b)} Transmission spectrum when illuminating the cavity with a white light LED (orange) and free space emission spectrum of the emitter (blue). The next higher TEM$_{00}$ mode ($\lambda \approx 689\,$nm) of the cavity is used for locking. \textbf{Inset:} Spectral profile of the locked cavity on the emitter, probed with a tunable laser, revealing spectral narrowing of a factor of 160 compared to free space emission. 
\textbf{c)} Measured spectral density $\delta_{\text{sp}}$ for the emitter in free-space configuration (blue) and calculated spectral density for the coupled emitter-cavity system (orange). A spectral enhancement of a factor of 100$\pm40$ induced by cavity funneling is obtained. \textbf{d)} Normalized second-order autocorrelation histogram of the coupled emitter-cavity system (blue) and corresponding fitted function (orange), which indicates single photon emission with $g^{(2)}(0)=0.3$. \textbf{e)} Pulsed lifetime measurement (start-stop histogram) of the emitter-cavity system using a pulsed laser (blue). An optical lifetime of $\tau_0 = 4.9 \pm 0.2\,$ns is extracted from a fitted exponential decay function (orange). \textbf{f)} Spectral fluorescence of the emitter cavity system for different cavity lengths $L_{\text{Cav}}$. The original cavity length $L_{\text{Cav}}\approx5.5\,$µm is tuned linearly by $\Delta L_{\text{Cav}}$. \textbf{g)} Finesse $\mathcal{F}$ of the coupled cavity-emitter system (orange) and the empty cavity (black) for different wavelengths $\lambda$. The values for the finesse are extracted from cavity transmission scans with a narrowband tunable laser. The coating-defined finesse (blue) is calculated from mirror transmission measurements provided by Laseroptik GmbH. \textbf{h)} Calculated losses $S$ introduced by the integration of the emitter (black) and corresponding fitted polynomial function (red). Losses around the ZPL of the emitter are within the measurement uncertainty, indicating the possibility of experiments with even higher finesse in the future.}
\label{Coupled_System}
\end{figure*}

\section{Cavity Integration}

The FPFC is designed for optimized detection of integrated SPEs while simultaneously allowing the observation of optomechanical interaction between the membrane and the intracavity field. 
A major challenge arises from the integration of the hBN host into the cavity mode, which inevitably leads to additional losses, caused e.g. by scattering. In order to be able to efficiently detect the SPEs fluorescence, the total optical losses $\mathcal{L}_{\text{opt}}= \frac{2L}{c} \kappa = \frac{2L}{c} \left(\kappa_S + \kappa_A + \kappa_o\right)$ need to be dominated by the out-coupling rate of the cavity mirror $\kappa_o$, compared to the loss rates given by scattering and absorption $\kappa_S$ and $\kappa_A$. The design finesse $\mathcal{F} =\frac{2\pi}{\mathcal{L}_{\text{opt}}}$ of the system ($\mathcal{F}_{\text{max}}\lesssim 4000$, $\mathcal{L}_{\text{opt}} \gtrsim 1550\,$ppm) is therefore chosen to be comparable to previous experiments with SPEs in solid-state materials \cite{Haeussler2021,Herrmann2024,Berghaus2025,Sachero2025} and reduced compared to previous optomechanical experiments \cite{Rochau2021,SanchezArribas2023}. As a result, the optomechanical resolution, which is linked to the cavities linewidth, is reduced as compared to state-of-the-art optomechanics experiments.\\
Minaturization allows however to boost the optomechanical coupling rate $g_{\text{om}}$ and the optical coupling rate between the cavity mode and the SPE $g_{\text{opt}}$ by reducing the optical mode Volume $V_m \approx \frac{\pi}{4}\omega_0^2L$ via microscopic curved mirror structures \cite{Maier2025b} (ROC$<57\,$µm) on the tip of an optical fiber. The cavity length $L$ is chosen small enough to observe the SPEs fluorescence and, at the same time, large enough to position the membrane within the cavity mode, boosting the optomechanical coupling rate $g_{\text{om}}$. In our system, cavity lengths below $L= 21\,$µm are routinely achieved in a MiM configuration, corresponding to an optical mode waist radius $\omega_0\lesssim1.2\,$µm and a modal volume $V_m \lesssim 24\,$µm$^3$. Assuming a natural linewidth of $\gamma_{\text{SPE}} = 60\,$MHz in a SPE in hBN \cite{Dietrich2018,Haeussler2021}, an optical coupling rate of $g_{\text{opt}}\approx 8.7\,$GHz could be realized. In principle, even shorter cavity lengths are possible but the minimum cavity length is ultimately limited by the need to avoid contact between the fiber mirror and the Si$_3$N$_4$ membrane. To precisely control the geometrical parameters of the curved mirror surface a combined fabrication of FIB milling and subsequent CO$_2$ laser smoothing is used \cite{Maier2025b}, which enables Gaussian mode matching, required to efficiently excite the cavity mode as well as collect the coupled fluorescence of the SPE. 
Miniaturization of the optical cavity simultaneously enables a reduction in the dimensions of the mechanical resonator, whose lateral size is ultimately limited by the optical mode waist radius $\omega_0\lesssim1.2\,$µm. Depending on the operating mode, either the Si$_3$N$_4$ membrane or the planar mirror can be tuned in lateral position with piezo driven actuators, enabling the system to act as a scanning cavity microscope \cite{Mader2015}. For more details on the FPFC systems and integration of suitable SPEs into the cavity, we refer to supplementary material section III.

\section{Cavity coupled single photon emitters in \NoCaseChange{h}BN flakes}
\label{Coupled_System}

 \begin{figure*}
    \centering
    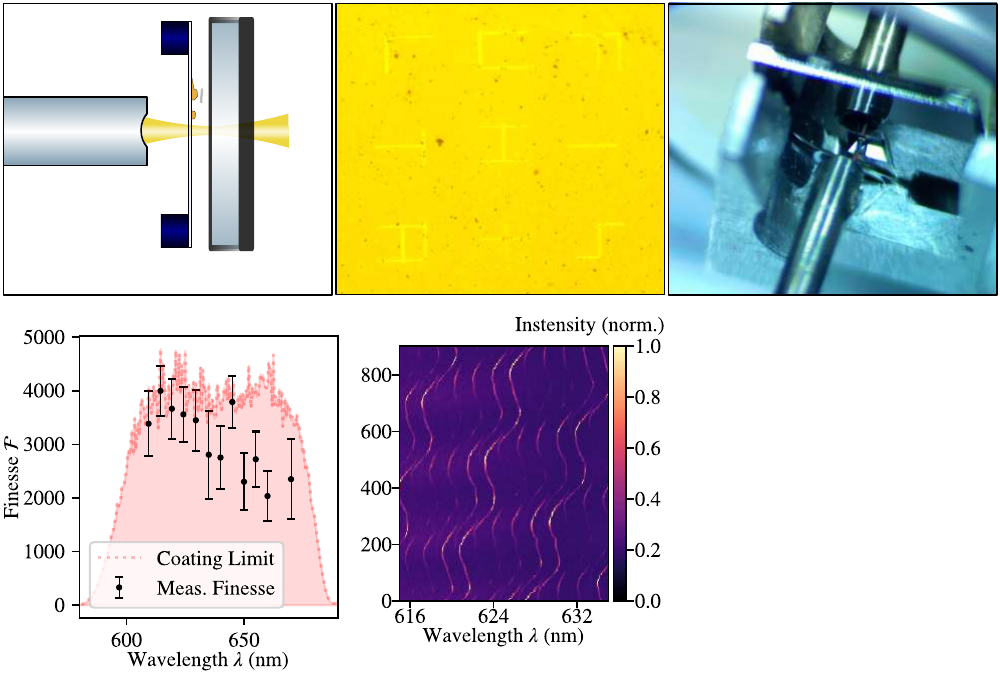
    \caption{Hybrid optomechanical system based on an open FPFC resonator with an optomechanical membrane and flakes of hBN. \textbf{a)} Schematic representation of the experimental configuration. A membrane with a multitude of flakes of hBN is integrated into a fiber resonator. The membrane can be positioned in lateral direction (xy) with piezo driven slick-stick actuators (Attocube Systems AG) and tuned along the optical axis (z) with a piezo assited linear stage. \textbf{b)} Light microscope image of the central region of a 30$\,$nm thick membrane equipped with platinum sputtered markers and drop-casted flakes of hBN. \textbf{c)} Observation microscope image of the experimental apparatus including a planar DBR mirror (I), the Si$_3$N$_4$-membrane on a Si frame (II), the cavity fiber (III) and a metal holder for free standing positioning of the membrane inside the micro-cavity (IV). \textbf{d)} Finesse of the system with optimized position of the membrane for different wavelengths (black) and coating designed limit (red). \textbf{e)} Cavity transmission when illuminated with a broadband LED for different membrane positions, from which the frequency pulling parameter $G_{\text{max}}\approx8.3\,\frac{\text{GHz}}{\text{nm}}$ and the cavity length $L_{\text{Cav}}\approx21\,$µm are extracted. \textbf{f)} Finesse of the cavity for different membrane positions (blue) and corresponding sinusoidal fit (red) at $\lambda = 624.5\,$nm. The periodically occurring losses are attributed to scattering in higher optical modes due to remaining misalignment between cavity mode axis and membrane \cite{SanchezArribas2023}. }
\label{Optomechanical Experiment}
\end{figure*}

Coupling the optical dipole of the SPE in a transferred flake to the optical mode of the FPFC results in a narrowed and enhanced emission compared to the free-space emission (fig. \ref{Coupled_System} a)). Since the emission linewidth of the coupled system is below the resolution limit of our spectrometer, the linewidth of the coupled emitter-cavity system is probed with a tunable laser (Dye ring laser system), while the cavity is stabilized with a second laser at the next higher TEM$_{00}$ mode (fig. \ref{Coupled_System} b)), yielding a locked linewidth of 8.9$\,$GHz. This linewidth corresponds to a spectral narrowing factor of $\approx$160 compared to the free space emission of the emitter. Tuning the length of the cavity allows us to probe and reconstruct the free space emission spectrum of the emitter (depicted normalized in blue in fig. \ref{Coupled_System} a)). We compare the free space emission of the emitter with the cavity coupled emission using the same optical setup and components, only flipping the mirror surface towards the objective. The spectral density $\delta$ can be calculated for both cases (free space emission and cavity coupled emission) when exciting the system off-resonantly at $\lambda = 532\,$nm, which yields a 100$\pm40\,$-fold enhanced spectral density $\delta_{\text{Cav}}$ (fig. \ref{Coupled_System} c)) in cavity configuration. This enhancement is caused by cavity funneling of the thermally broadened emission (bad emitter regime) into the cavity mode \cite{Grange2015, Janitz2020, Husel2024}.
Pulsed lifetime measurements indicate an optical lifetime of the coupled system of $\tau_0=4.9 \pm 0.2\,$ns which is comparable to the free space lifetime. Due to the operation at room temperature the emission is thermally broadened resulting in a reduced quality factor $Q_\text{hBN}$ \cite{Albrecht2013}. This quality factor is lower compared to the cavity quality factor $Q_\text{Cav}$ (bad emitter regime), which makes observation of Purcell enhanced emission harder compared to cryogenic systems \cite{Feuchtmayr2023,Sachero2025,Berghaus2025}.                                                                                                                                                                                                                                                                                                                                                                                                                                                                                                                                                                                                                                                                                                                                                                                                                                                                                                                                                                                                                                                                                                                                                                                                                                                                                                                                                                                                                                                                                                                                                                                                                                                                                                                                                                                                                                                                                                                                                                                                                                                                                                                                                                                                                                                                                                                                                                                                                                                                                                                                                                                                                                                                                                                                                                                                                                                                                                                                                                                                                                             Second-order auto correlation measurements (fig. \ref{Coupled_System} d)) reveal the single-photon emitting character of the cavity coupled emission after fitting an exponential function with $g^2(0)=0.3$. To determine the impact of the host material on the system, we conduct finesse measurements and compare the coupled system with the empty cavity and the coating designed finesse as a reference (fig. \ref{Coupled_System} g)). For more details see supplementary material section IV. We directly compare finesse values for corresponding wavelengths and fit a sixth order polynomial function into each dataset (empty cavity and coupled system) to determine the optical losses for different wavelengths (depicted in fig. \ref{Coupled_System}h)). We are not able to resolve optical losses introduced by the flake structure $\mathcal{L}_{\text{hBN}}$ at the ZPL wavelength of the emitter beyond the error of our measurements, indicating optical losses are dominanted by the outcoupling rate $\kappa$ defined by the mirror coatings. The vanishing contrast in finesse between the coupled system and the empty cavity indicates that the system is no more limited in its quality factor $Q_o$ by the introduction of the host material, which can be attributed to its manipulated topography.
To further investigate the effects of the integrated material on the optical properties of the system we probe the dispersion relation of the cavity-emitter system. Linearly tuning the cavity length of the coupled system
while acquiring spectra at each step, yields a linear correlation
between the resonant frequency of the cavity-emitter
system and the length detuning (as seen in fig. \ref{Coupled_System} f)), attributed to the reduced lateral dimension of the flake compared to membranes \cite{Haeussler2019, Koerber2023}.

\begin{figure*}
    \centering
    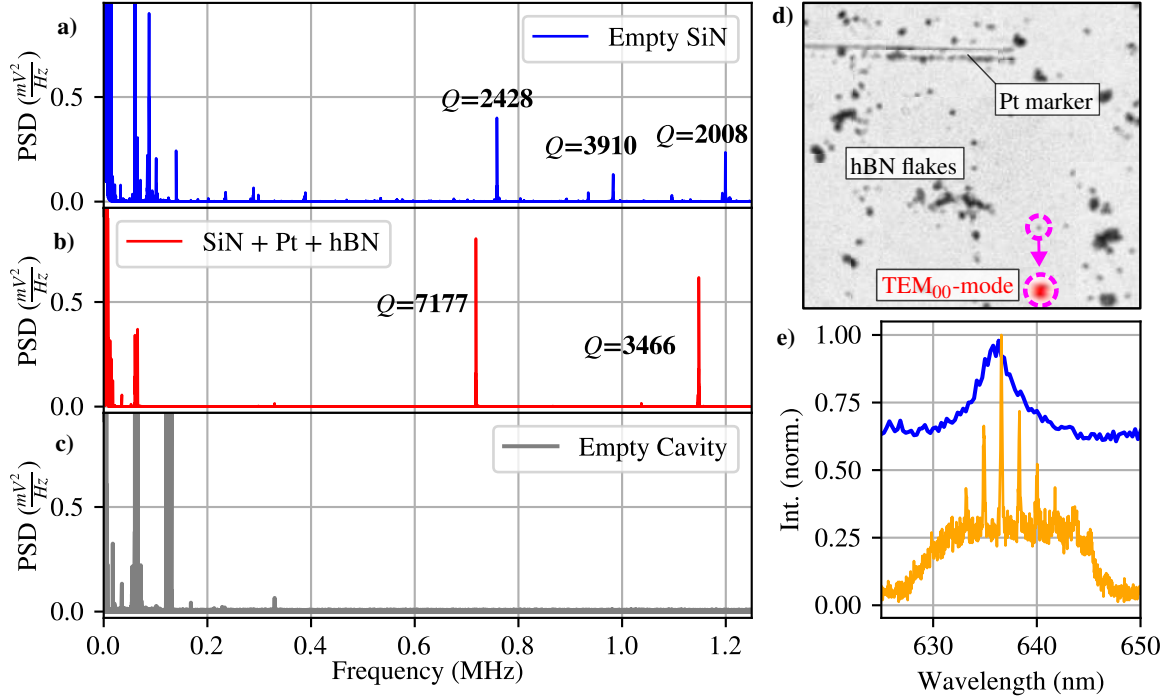
    \caption{Observation of mechanical modes and cavity coupled fluoroescence of a SPE in hBn in a MiM-FPFC. \textbf{a)} Power spectral density spectrum for cavity with an integrated Si$_3$N$_4$-membrane.The bare membrane is characterized as a reference prior to the incorporation of hBN flakes and subsequent fabrication steps. \textbf{b)} Power spectral density spectrum for the same membrane as in a), but equipped with Pt markers and hBN flakes. \textbf{c)} Power spectral density spectrum for an empty cavity without membrane or hBN used as reference. \textbf{d)} Microscope image through the planar side of the cavity using a wavelength at which the planar mirror is partially translucent ($\lambda=405\,$nm). A hBN flake hosting an SPE is encircled in purple. When the cavity is probed with a narrow laser, its optical mode waist can be located on the membrane (depicted in red). Subsequent movement of the SPE into the mode allows coupling to the cavity mode. \textbf{e)} Cavity coupled emission (orange) of the SPE marked in e) when excited with a $\lambda=532\,$nm laser. The free space emission is depicted in blue.}
\label{OptomechanicsHBN}
\end{figure*}

\section{Hybrid opto-mechanical system with individual SPE\NoCaseChange{s} in \NoCaseChange{h}BN}

The cavity is modified with a high-stress Si$_3$N$_4$-membrane (Norcada) in a MiM configuration, as depicted in fig. \ref{Optomechanical Experiment}a),c). The membrane is positioned with slip-stick actuators (Attocube Systems), while the planar mirror is mounted in a fixed position. The  Si$_3$N$_4$ membrane is equipped with platinum markers fabricated in an electron microscope (supplementary material section V) and hBN flakes in solution are drop-casted onto the surface. Pre-characterization of the resulting hybrid system includes light microscope images (fig. \ref{Optomechanical Experiment}b)), confocal measurements and subsequent AFM manipulation of individual flakes and clusters (supplementary material section V). After integration of the membrane into the cavity, the finesse of the MiM-FPFC is found to be within the coating-designed limit (fig. \ref{Optomechanical Experiment} d)). When varying the membrane position and probing the system with a broadband LED, modulation of individual transmission modes can be observed. From the measured mode shifts the frequency pulling factor $G(z)$ can be determined yielding $G_{\text{max}}\approx 8.3\,$GHz at a cavity length of $L_{\text{Cav}}\approx21\,$µm (fig. \ref{Optomechanical Experiment} e)). Probing the system with a narrowband laser ($\lambda=624.5\,$nm) and tuning the membrane position reveals a position-dependent finesse $\mathcal{F}(z)$ (fig. \ref{Optomechanical Experiment} f)) related to scattering into higher optical modes due to residual misalignment between cavity mode axis and membrane \cite{SanchezArribas2023}. \\
In order to probe for mechanical motion of the membrane inside the resonator, the system is evacuated ($p\approx 2\cdot 10^{-3}\,$mbar) and the cavity length is stabilized with a Pound-Drever-Hall (PDH) cavity lock. The system is then probed with a narrow-band tunable laser and time traces of the transmitted signal are recorded with an avalanche photo diode (APD). Calculation of the spectral power density reveals mechanical membrane resonances exceeding $\Omega_m > 1\,$MHz (fig. \ref{OptomechanicsHBN}a),b
)), which is in accordance with previous measurements of membranes with a similar geometry \cite{Thompson2008,Jayich2008,FlowersJacobs2012}. The fabrication of marker patterns and the integration of hBN flakes introduces a shift in frequency while keeping the quality factor of the modes in a comparable range. It is noted that the mechanical quality factor $Q_{\text{$Si_3N_4$}}$ is primarily limited in both configurations by remaining air pressure \cite{Bao2002,Chakram2014} and contact of the silicon frame with the metal holder \cite{Chakram2014} rather than by introduction or manipulation of hBN flakes or Pt markers.\\
In a next step, a flip mirror allows the objective to be used in a wide field microscope configuration to observe the membrane inside the closed cavity in planar mirror transmission. With this technique, individual flakes can be positioned inside the optical mode of the cavity. Confocal pre-characterisation, AFM manipulation and orientation with help of Pt markers (fig \ref{OptomechanicsHBN}d)) allows deterministic pre-selection of individual SPEs. Excitation with an off-resonant laser at $\lambda=532\,$nm reveals cavity coupled emission from an individual SPE (fig. \ref{OptomechanicsHBN}e)). It is noted that the signal to noise ratio is reduced compared to the SPE described in the previous section. This is attributed to the increased cavity length ($L\approx21\,$µm) and ROC ($ROC\approx 57\,$µm) of the curved cavity structure, introduced to avoid contact between cavity fiber and membrane. Future improvements include spectrally narrower SPEs, e.g. by further sample treatment methods like thermal annealing, reducing thermal broadening by cooling \cite{Purdy2012,Ruelle2022a} and the use of a closed-loop piezo systems to control the distance between cavity fiber and membrane enabling shorter cavity lengths $L_{\text{cav}}$.

\section{Conclusion and Outlook}


In our work, we explore the capabilities of SPEs in hBN flakes as source for SPE-optomechanical system in a Si$_3$N$_4$-membrane-in-the-middle cavity configuration. Establishing a toolset of manipulation techniques to decluster and extract individual hBN flakes containing SPEs from commercially available hBN solution enables the deterministic transfer into a FPFC. The thickness of the extracted flakes reaches below 27$\,$nm comparable to membranes originating from mechanical exfoliation \cite{SanchezArribas2023}. Scattering effects are reduced by manipulating the flakes topography enabling the coupling of a spectrally narrow single photon emitter to the mode of an open FPFC with increased optical quality factors $Q_o$ and spectrally enhanced emission by a factor of 100. The lateral extent of the hBN flakes remain small, about three-times smaller as compared to the optical mode waist diameter, not suitable for direct implementation as optomechanical system. However, membranes with larger lateral extend could be investigated by our manipulation techniques in the future. As an alternative route we introduce highly strained Si$_3$N$_4$-membranes into the cavity in a membrane-in-the-middle-configuration. We reproduce the preparation and manipulation of hBN flakes on the Si$_3$N$_4$-membrane and observe mechanical eigenmodes of the membrane above $\Omega_m = 1\,$MHz. The mechanical quality factor $Q_m$ remains similar to an untreated Si$_3$N$_4$-membrane, limited by background pressure and frame contact. Both limitations could be improved in future steps to achieve mechanical quality factors above $Q_m>10^5$ reported with similar membranes \cite{Rochau2021}. We achieve optical coupling of the cavity mode to a SPE inside the manipulated hBN flake positioned on the Si$_3$N$_4$-membrane. Reduced signal of the cavity coupled emitter compared to the coupling of an emitter into an empty FPFC is attributed to a higher cavity length and ROC of the fiber structure, which were introduced to prevent contact between fiber and
 Si$_3$N$_4$-membrane. The quality factor $Q_o= 5\cdot10^4$ measured with an hBN flake on the planar mirror can be extrapolated to the membrane-in-the-middle configuration. The frequency pulling parameter $G=8.3\,\frac{\text{GHz}}{\text{nm}}$ is defined by the Si$_3$N$_4$ membrane and would result in a single photon coupling rate of $\frac{g_0}{2\pi} \approx 200\,$Hz but could exceed $\frac{g_0}{2\pi}>10^4\,$Hz for tailored hBN flakes with an effective mass $m_{\text{eff}}$ in the range of few femtograms. \\
The hybrid approach could benefit from the availability of narrow spectral lines up to room temperature under resonant excitation schemes in the hBN flakes \cite{White2021,Hoese2020,Koch2024}. Previously reported SPEs susceptible to mechanical deformation \cite{Shaik2022, Grosso2017, Mendelson2020} could be investigated in the hybrid optomechanics platform \cite{SanchezArribas2023}. Spin defects have been reported in hBN \cite{Liu2022,Gottscholl2021,Gao2022,Stern2022,Guo2023,Stern2024, Wu2025} and recently also spin signatures in mechanical isolated emitters have been shown \cite{2026}.\\
Our work paves the way to construct complex quantum systems by combining optomechanics with SPEs and potentially with spin states \cite{Abdi2019}. Ultimately, such hybrid spin-optomechanics systems could enable to test spin-mechanical schemes \cite{Wang2020} with the ability to engineer spin-motion interaction \cite{Barzanjeh2022}. The high level of complexity in the system, arising from coupling multiple and radically different degrees of freedom, leads to potential for a broad range of applications. Quantum Information could be exchanged and processed between photons, phonons and stationary qubits in hBN. Ultra-sensitive force detection \cite{Moser2013,Muschik2014} as well as access to quantum simulation of 2D spin systems \cite{Cai2013} are among potential applications.


\begin{acknowledgments}
The authors gratefully acknowledge support of the Baden-Wuerttemberg Stiftung gGmbH in project AmbientCoherentQE. The AFM was funded by the DFG. We thank Prof. Dr. Kay Gottschalk for support. Prof. Dr. Christine Kranz, Dr. Gregor Neusser and the Focused Ion Beam Center UUlm are acknowledged for their scientific support during FIB
milling. We thank Prof. Dr. Benjamin Stickler for fruitful discussions. We thank Jens Fuhrmann and Prof. Dr. Fedor Jelezko for sample annealing. Manuel Mundszinger is acknowledged for support during SEM imaging. Measurements were conducted among others with the Qudi software suite \cite{Binder2017}. AFM scans were evaluated among others with the open source software Gwyddion \cite{Gwyddion}.
\end{acknowledgments}

\section*{Data Availability Statement}

The data that support the findings of this study are available from the corresponding author upon reasonable request.

\bibliography{aipsamp}

\end{document}